\documentclass{turintech-v2}

\usepackage[utf8]{inputenc}
\usepackage[T1]{fontenc}
\usepackage{hyperref}
\usepackage{cleveref}
\usepackage{url}

\makeatletter
\renewcommand\addtolist[5][]{%
  \begingroup
    \if\relax#3\relax\def\sep{}\else\def\sep{#5}\fi
    \let\protect\@unexpandable@protect 
    \xdef#3{\expandafter{#3}\sep #4[#1]{#2}}%
  \endgroup
}
\makeatother

\newcommand\authorlist{}
\newcommand\authorformat[2][]{\textbf{#2}$^{#1}$}
\renewcommand\author[2][]{\addtolist[#1]{#2}{\authorlist}{\authorformat}{, }}

\newcommand\affiliationlist{}
\newcommand\affiliationformat[2][]{$^{#1}$#2}
\newcommand\affiliation[2][]{\addtolist[#1]{#2}{\affiliationlist}{\affiliationformat}{~~}}

\newcommand\contributionlist{}

\newcommand\metadatalist{}
\makeatletter
\newcommand\addmetadata[1]{%
  \ifx\metadatalist\empty
    \gdef\metadatalist{#1}%
  \else
    \g@addto@macro\metadatalist{\par\vspace{0.15em}#1}%
  \fi
}
\makeatother
\renewcommand\date[1]{\addmetadata{\textbf{Date:} #1}}

\makeatletter

\newcommand\formatcodeurls[1]{%
  \begingroup
  \setlength{\parindent}{0pt}%
  \setlength{\parskip}{0.1em}%
  \def\codesep{}%
  \@for\next:=#1\do{%
    \codesep\url{\next}%
    \def\codesep{\\}%
  }%
  \endgroup
}
\makeatother

\newcommand\projecturl[1]{\addmetadata{\textbf{Project:} \url{#1}}}

\usepackage{amsfonts}
\usepackage{amsmath}
\usepackage{amssymb}
\usepackage{amsthm}
\usepackage{mathtools}
\usepackage{nicefrac}

\newtheorem{definition}{Definition}

\usepackage{nicefrac}
\usepackage{tikz}
\usepackage{pgfplots}
\pgfplotsset{compat=1.16}
\usepackage{xcolor}

\usepackage{algorithm}

\usepackage{booktabs}
\usepackage{multirow}
\usepackage{multicol}
\usepackage{array}

\usepackage{listings}
\usepackage{tcolorbox}
\usepackage{xspace}
\usepackage[frozencache=true,cachedir=minted-cache]{minted}

\graphicspath{{images/}{../paper/}}

\newcommand{\Artemis}{\textsc{Artemis}\xspace}

\title{Evolving Excellence: Automated Optimization of LLM-based Agents}

\author[1]{Paul Brookes}
\author[1]{Vardan Voskanyan}
\author[1, 6]{Rafail Giavrimis}
\author[1]{Matthew Truscott}
\author[1]{Mina Ilieva}
\author[1]{Chrystalla Pavlou}
\author[1]{Alexandru Staicu}
\author[1]{Manal Adham}
\author[1]{Will Evers-Hood}
\author[1, 2]{Jingzhi Gong}
\author[1, 5]{Kejia Zhang}
\author[1]{Matvey Fedoseev}
\author[1]{Vishal Sharma}
\author[6]{Roman Bauer}
\author[7]{Zheng Wang}
\author[3]{Hema Nair}
\author[4]{Wei Jie}
\author[7]{Tianhua Xu}
\author[5]{Aurora Constantin}
\author[8]{Carmine Ventre}
\author[1]{Leslie Kanthan}
\author[1]{Michail Basios}

\affiliation[1]{TurinTech AI, London, UK}
\affiliation[2]{University of Leeds, Leeds, UK}
\affiliation[3]{City, University of London, London, UK}
\affiliation[4]{University of West London, London, UK}
\affiliation[5]{University of Edinburgh, Edinburgh, UK}
\affiliation[6]{
University of Surrey, Guildford, United Kingdom}
\affiliation[7]{University of Warwick, Coventry, UK}
\affiliation[8]{King's College London, London, UK}

\begin{document}

\abstract{ Agentic AI systems built on large language models (LLMs) offer
significant potential for automating complex workflows, from software
development to customer support. However, LLM agents often underperform due to
suboptimal configurations; poorly tuned prompts, tool descriptions, and
parameters that typically require weeks of manual refinement. Existing
optimization methods either are too complex for general use or treat components in isolation, missing critical interdependencies.

We present \Artemis, a no-code evolutionary optimization platform that jointly
optimizes agent configurations through semantically-aware genetic operators.
Given only a benchmark script and natural language goals, \Artemis
automatically discovers configurable components, extracts performance signals
from execution logs, and evolves configurations without requiring architectural
modifications.

We evaluate \Artemis on four representative agent systems: the \emph{ALE
Agent} for competitive programming on AtCoder Heuristic Contest, achieving a
\textbf{\textcolor{blue!70!black}{$13.6\%$ improvement}} in acceptance rate;
the \emph{Mini-SWE Agent} for code optimization on SWE-Perf, with a
statistically significant \textbf{\textcolor{blue!70!black}{10.1\% performance
gain}}; and the \emph{CrewAI Agent} for cost and mathematical reasoning on Math
Odyssey, achieving a statistically significant \textbf{\textcolor{blue!70!black}{$39.1\%$ reduction}} in the number
of tokens required for evaluation. We also evaluate the
\emph{MathTales-Teacher Agent} powered by a smaller open-source model
(Qwen2.5-7B) on GSM8K primary-level mathematics problems, achieving a
\textbf{\textcolor{blue!70!black}{37.3\% accuracy improvement}} and demonstrating that \Artemis can optimize
agents based on both commercial and local models. These initial results
indicate that \Artemis can deliver substantial improvements for agents with
optimization potential, reducing configuration tuning time while uncovering
non-obvious optimizations.  Our results show that automated agent optimization
through \Artemis is practical and effective, making sophisticated optimization
accessible to practitioners \emph{without requiring deep expertise} in
evolutionary algorithms.}

\maketitle


\newpage

\section{Introduction}\label{sec:intro} 
Large Language Model (LLM) agents are
being rapidly deployed across domains ranging from software
engineering~\cite{yang2024sweagent} to scientific
discovery~\cite{ren2025towards}. Despite their growing promise, the potential of these systems is often hindered by fragile and suboptimal configurations. While increasingly sophisticated architectures like
\textbf{ReAct}~\cite{yao2022react}, \textbf{Tree of
Thoughts}~\cite{yao2023tree}, and \textbf{Reflexion}~\cite{shinn2023reflexion} have been developed, practitioners frequently find that even minor misconfigurations in prompts, tool descriptions, or execution parameters can cause {substantial} drops in performance~\cite{mast2024}. As we will show later,
systematic optimization can improve agent performance by \textbf{9.3--13.6\%}, yet many practitioners rely on manual trial-and-error approaches that are both time-consuming and brittle.

The configuration challenge arises because modern LLM agents are not monolithic systems but \textbf{multi-component pipelines}. A typical agent coordinates system prompts, tool documentation, few-shot examples, and retry strategies, all of which interact in subtle ways. The resulting configuration space is \emph{high-dimensional and heterogeneous}, encompassing natural language, discrete design choices, and continuous parameters. {In practice, manual tuning in such settings can require extensive trial-and-error and still produce configurations that do not reliably generalize across tasks or deployment environments.}

There are several fundamental sources of this difficulty. First,
\textbf{non-trivial component interactions} mean that optimizing prompts in
isolation, without considering specific tool configurations and other
components, results in lower performance. Second, \textbf{non-differentiable objectives} prevent the use of standard optimization techniques such as gradient descent. Third, \textbf{expensive evaluation} makes exhaustive search infeasible, as each candidate may require minutes to hours of benchmark execution. Fourth, \textbf{task-specific optima} rarely transfer across domains; prompts optimized for one task (e.g., mathematics) may fail on others (e.g., code generation). Finally, \textbf{sensitivity to perturbations} makes
manual tuning fragile, as even slight wording changes can cause dramatic performance shifts.

Existing optimization methods provide only partial relief. Prompt optimization
techniques like \textbf{APE}~\cite{zhou2022ape} and
\textbf{PromptBreeder}~\cite{fernando2023promptbreeder} operate in isolation,
while tool-focused approaches overlook the critical interactions between
components. Although recent work begins to explore broader agent and workflow
optimization~\cite{adas2024,aflow2024}, these methods typically target specific
architectural assumptions or workflow representations. To date, few existing systems provide unified, end-to-end frameworks for optimizing complete agent pipelines while remaining agnostic to the underlying agent implementation. Moreover, current methods typically ignore
the wealth of information available in execution logs and benchmark feedback,
leaving untapped opportunities for systematic improvement. Current manual and
semi-automated tuning methods suffer from three critical limitations.
\textbf{Manual tuning} relies on expert intuition and trial-and-error, making
it labor-intensive (often requiring substantial engineering effort per
agent), non-reproducible (different engineers produce different configurations), and
non-scalable (expertise does not readily transfer across domains or team
members). \textbf{Semi-automated approaches} like grid search or random
sampling can explore parameter spaces but lack semantic understanding; they cannot reason about why a prompt works or intelligently combine successful patterns. 

{Existing \textbf{LLM-based optimizers} (e.g.,
DSPy~\cite{dspy2024lnbip}) focus primarily on prompt refinement rather than full agent pipelines, and their reported effectiveness varies widely across tasks, suggesting that current optimization strategies still leave substantial room for improvement.}

In this paper, we demonstrate how one can apply
\textbf{\Artemis}~\cite{artemis2025}, a general-purpose evolutionary
optimization platform, for automated agent configuration tuning. Rather than restricting optimization to a single component (such as a system prompt), \Artemis treats agents as \textbf{black boxes}, requiring no
architectural modifications, and can jointly optimize multiple
configurable components, both textual and parametric, while capturing their
interdependencies. The system leverages benchmark outcomes and execution logs
as feedback, applying \emph{semantically-aware} mutation and crossover operators specifically tailored for natural language components. In practice, \Artemis automates most stages of the tuning workflow and is usable with only limited coding effort, making it accessible to practitioners without specialized evolutionary-optimization expertise. Users specify optimization goals declaratively (often in natural language) and configure how benchmark performance is translated into fitness scores, after which the platform orchestrates the search over candidate configurations.

We evaluate \Artemis across four distinct and representative applications:
(1)~the \textbf{ALE Agent}~\cite{alebench2025} on the \textbf{AtCoder Heuristic
Contest}, tackling competitive programming problems; (2)~the \textbf{Mini-SWE
Agent}~\cite{miniSWEagent2024} on the \textbf{SWE-Perf} benchmark~\cite{he2025swe}, optimizing code performance; and (3)~an agent built using the \textbf{CrewAI} framework to work on the \textbf{Math Odyssey} benchmark~\cite{mathodyssey2025}, handling mathematical reasoning. (4)~the \textbf{MathTales-Teacher Agent} powered by Qwen2.5-7B on the \textbf{GSM8K} benchmark~\cite{cobbe2021gsm8k}, solving primary-level mathematics problems. These domains span fundamentally different challenges: algorithmic reasoning, performance optimization, and mathematical discourse, allowing us to assess whether low-code, black-box optimization generalizes across agent architectures and task characteristics.

Our results show consistent improvements for agents with optimization
potential: the \textbf{ALE Agent} improved by \textbf{13.6\%} in competitive
programming acceptance rate, the \textbf{Mini-SWE Agent} gained \textbf{10.1\%} in code optimization tasks, while the \textbf{CrewAI Agent} showed a significant \textbf{39.1\% decrease} in execution cost, at the price of a small but statistically significant drop in correctness. The \textbf{MathTales-Teacher Agent} achieved \textbf{37.3\%} accuracy improvement on primary-level mathematics, demonstrating that \Artemis effectively optimizes agents based on smaller open-source models. These findings suggest that while well-tuned agents may offer limited room for further optimization, systematic tuning through \Artemis can substantially enhance under-optimized systems.

Beyond reporting raw performance numbers, our study addresses several key
questions: \begin{itemize} \item \textbf{Performance:} Can evolutionary
optimization significantly improve the overall performance of different
LLM-based agents in terms of accuracy and efficiency?  \item
\textbf{Generalization:} Does a {low-code, black-box} optimization
framework transfer across varied domains?  \item \textbf{Limitations:} Under
what conditions does automated optimization provide little or no benefit?
\end{itemize}

The main contributions of this work are:
{\begin{enumerate}[leftmargin=*,
itemsep=2pt, topsep=2pt]
\item We introduce the \Artemis platform as a tool for
optimizing the textual and parametric components of LLM agent pipelines without requiring architectural modifications to the agent implementation or
specialized evolutionary-algorithm expertise.
\item \Artemis incorporates novel
mutation and crossover operators specifically designed for natural language components, leveraging LLM ensembles to perform intelligent mutations and crossovers that maintain validity while exploring the configuration space.
\item We conduct systematic experiments across four agentic systems and benchmarks (competitive programming, code optimization,
mathematical reasoning), with rigorous statistical validation including
confidence intervals and non-parametric tests.  
\item Our analysis reveals key
factors influencing optimization success, including the relationship between
initial configuration quality and improvement potential, task characteristics
that affect success, and computational cost-benefit trade-offs for production
deployment.  
\end{enumerate}}

The remainder of this paper is organized as follows.
Section~\ref{sec:related}
reviews related work in prompt optimization, agent design, and evolutionary
methods. Section~\ref{sec:problem} formalizes automated agent optimization as a
mixed-type evolutionary problem. Section~\ref{sec:artemis} introduces the
\Artemis platform and its optimization workflow. Section~\ref{sec:setup}
details our experimental setup and evaluation methodology, while
Section~\ref{sec:results} presents empirical results addressing our research
questions. Section~\ref{sec:discussion} discusses broader implications and
limitations. Finally, Section~\ref{sec:conclusion} concludes with future
research directions.

\section{Related Work}\label{sec:related}

The optimization of large language model (LLM) agents has become a
rapidly expanding area of research. Existing work broadly clusters into
four paradigms: prompt engineering, workflow optimization, multi-agent system
design, and holistic agent optimization.

\textbf{Prompt Engineering Approaches.}
Prompt optimization has evolved from manual engineering to automated methods. A comprehensive survey
\cite{ijcai2024promptsurvey} identifies over 41 distinct prompting techniques across diverse application areas,
highlighting both the vastness of the search space and the lack of any universally optimal strategy; effectiveness remains highly dependent on task and system context.

Early automated approaches, such as APE \cite{zhou2022ape}, demonstrated that LLMs could improve their own prompts, outperforming human-designed prompts on 19 out of 24 NLP tasks. PromptBreeder \cite{fernando2023promptbreeder} extended this idea by using evolutionary algorithms for self-referential prompt evolution, establishing an early application of evolutionary search to natural language optimization. An industrial study \cite{gong2025tuning} complements these academic results by leveraging meta-prompting strategies in production-scale code optimization pipelines and integrating automated prompt tuning into real-world workflows.

Recent studies highlight both the strengths and limitations of automated
prompt optimization. For example, the DsPy framework analysis \cite{dspy2024lnbip} reports large gains for prompt-evaluation benchmarks (46.2\%--64\% accuracy) but only modest improvements across other tasks.
This variability indicates that prompt-level automation can be highly
effective in specific settings but does not transfer uniformly across domains.

\textbf{Evolutionary and Genetic Algorithms.}
A growing body of work combines evolutionary algorithms with LLM-based generation
to improve program synthesis, algorithm discovery, and automated repair.
ShinkaEvolve \cite{lange2025shinkaevolve} applies an island-model evolutionary
algorithm with LLM-driven mutation and crossover, novelty filtering via embedding
similarity, and adaptive ensemble control to improve sample efficiency.  
GEPA \cite{agrawal2025gepa} introduces Pareto-based evolutionary prompt optimization
using reflective natural language feedback, achieving performance competitive with
reinforcement-learning approaches without model weight updates.  
AlphaEvolve \cite{novikov2025alphaevolve} implements a closed-loop LLM generation,
execution, and verification pipeline that has produced novel algorithmic
improvements on scientific benchmarks.  
Related work explores hybrid evolutionary approaches using benchmark-driven
fitness and semantic evaluation strategies
\cite{van2024llamea, hemberg2024evolving, morris2024llm, pinna2025exploring}.
Collectively, these studies show that evolutionary search is effective for
optimizing non-differentiable objectives when paired with execution-based or
semantic feedback.

\textbf{Workflow and Architectural Optimization.}
A parallel research line focuses on optimizing agent workflows and internal
architectures.  
ADAS \cite{adas2024} represents agent pipelines as executable code structures and
optimizes their composition but relies on relatively limited experience encoding
for search guidance.  
AFlow \cite{aflow2024} extends this approach via named-node workflow abstractions
and Monte Carlo Tree Search (MCTS), enabling more expressive architectural search.

Domain-specialized workflow optimization has achieved strong results.
AlphaCodium \cite{alphaCodium2024} departs from prompt-only approaches in favor of
``flow engineering'' driven by multi-stage test-based refinement, improving GPT-4 performance on the CodeContests benchmark from 19\% to 44\%.
This outcome illustrates that tightly domain-aligned workflow optimization can substantially outperform generic prompting within narrowly scoped benchmark settings.

\textbf{Multi-Agent Systems and Failure Analysis.}
Despite theoretical advantages, multi-agent systems frequently underperform
relative to single-agent baselines.
The MAST taxonomy \cite{mast2024} provides the first empirically grounded analysis
of failures in Multi-Agent LLM Systems, identifying 14 failure modes grouped into
design/specification failures (41.77\%), coordination/communication failures
(36.94\%), and verification/termination failures (21.30\%). These results highlight the fragility of agent coordination and motivate the need for systematic configuration analysis and optimization.

\textbf{Agent Reasoning Frameworks.}
Modern LLM agents increasingly incorporate structured reasoning systems.
ReAct \cite{yao2022react} interleaves reasoning traces with tool-based actions,
achieving approximately 34\% absolute improvement on ALFWorld.
Tree of Thoughts \cite{yao2023tree} enables explicit exploration of reasoning paths,
improving performance on the Game of 24 benchmark from 4\% to 74\%.
Reflexion \cite{shinn2023reflexion} introduces memory-based self-feedback without
weight updates, reporting pass@1 rates exceeding 90\% on selected benchmarks.
Together, these frameworks demonstrate the rising complexity of agent
reasoning and further motivate automated methods for tuning agent configurations.

\textbf{Positioning Artemis.}
Several limitations remain evident:

\begin{itemize}
\item Most existing approaches optimize isolated components, such as prompts or
      workflows, limiting gains achievable through joint multi-component tuning.
\item Many methods require source-code modification or internal API access,
      constraining their application to closed or proprietary systems.
\item Optimization strategies frequently rely on narrow heuristics or tree-based
      searches with limited exploration diversity relative to evolutionary methods.
\item Empirical evidence across benchmarks indicates that generalization
      across agent types and domains remains inconsistent and task-dependent.
\end{itemize}

Our primary contribution is the design and evaluation of a general-purpose optimization platform that integrates ideas such as semantic prompt evolution, evolutionary improvement, and benchmark-driven fitness evaluation into a cohesive system for practical agent tuning.

{\Artemis is architecturally agnostic, operating through input--output
interaction and execution logs rather than internal code modifications, enabling deployment across heterogeneous agent frameworks with minimal integration effort.}

At the same time, the platform maintains transparency by exposing candidate mutations, fitness scores, and lineage information to users, supporting optional human-in-the-loop inspection and intervention during optimization.

\textbf{Comparative Analysis.}
Table~\ref{tab:framework_comparison} summarizes representative optimization
frameworks across optimization scope, generality, architectural agnosticism,
semantic awareness, and scalability.

\begin{table}[H]
\centering
\caption{Comparative analysis of LLM agent optimization frameworks}
\label{tab:framework_comparison}
\setlength{\tabcolsep}{4pt}
\begin{tabular}{lccccc}
\toprule
\textbf{Framework} & \textbf{Scope} & \textbf{Generality} & 
{\textbf{Arch.-agnostic}} & \textbf{Semantic} & \textbf{Scalable} \\
\midrule
APE            & Prompts  & High   & Yes & Limited & High   \\
PromptBreeder  & Prompts  & High   & Yes & Medium  & Medium \\
ADAS           & Workflow & Medium & No  & No      & Medium \\
AFlow          & Workflow & Medium & No  & No      & High   \\
AlphaCodium    & {Workflow (domain)} & {Low} & No  & {Medium} & Medium \\
GEPA           & Prompts  & High   & Yes & Medium  & Medium \\
ShinkaEvolve   & Code     & Medium & No  & Yes     & Low    \\
\midrule
\textbf{Artemis} & \textbf{Full agent} & \textbf{High} & \textbf{Yes} & \textbf{High} & \textbf{Medium} \\
\bottomrule
\end{tabular}
\end{table}

\section{Problem Formulation}\label{sec:problem}

We formalize agent optimization as evolutionary search over mixed-type configuration spaces, providing the mathematical foundation for the \Artemis approach.

\begin{definition}[Agent Configuration]
An LLM agent's configuration $\mathcal{C} = (P, T, M, \Theta)$ consists of:
\begin{itemize}[itemsep=2pt, topsep=0pt]
    \item $\mathcal{P} = \{p_1, ..., p_n\}$: Natural language prompts (system, user, assistant templates)
    \item $T = \{t_1, ..., t_m\}$: Tool descriptions, error messages, and usage instructions
    \item $M = \{m_1, ..., m_k\}$: Model assignments and routing decisions (discrete choices)
    \item $\Theta = \{\theta_1, ..., \theta_l\}$: Continuous parameters (temperature, thresholds, timeouts)
\end{itemize}
\end{definition}

The configuration space $\mathcal{S} = \mathcal{P} \times \mathcal{T} \times \mathcal{M} \times \mathbb{R}^l$ is thus a product space combining infinite-dimensional spaces of natural language ($\mathcal{P}, \mathcal{T}$), discrete model selections ($\mathcal{M}$), and continuous parameters ($\mathbb{R}^l$). This mixed-type space presents unique challenges: discontinuous performance landscapes, semantic validity constraints, and expensive black-box evaluation.

Given an agent $\mathbf{A}$ and benchmark $\mathbf{B}$, we seek the optimal configuration:
\begin{equation}
\mathcal{C}^* = \arg\max_{\mathcal{C} \in \mathcal{S}} f(\mathbf{A}; \mathcal{C}, \mathbf{B})
\end{equation}
where the fitness function $f: \mathcal{A} \times \mathcal{S} \times \mathcal{B} \rightarrow \mathbb{R}$ is derived by instantiating agent $A$ with configuration $\mathcal{C}$ and executing benchmark $B$ to obtain performance metrics. Importantly, Artemis does not require explicit mathematical aggregation formulas; instead, users provide natural language optimization goals (e.g., ``maximize accuracy while maintaining reasonable latency''), and the platform uses LLM-based semantic understanding to evaluate fitness based on benchmark results. In our experiments, we optimize the acceptance rate (ALE Agent), the performance score (Mini-SWE Agent), and cost and accuracy (CrewAI Agent).

The agent-optimization problem exhibits several characteristics that make evolutionary algorithms particularly well-suited. \textbf{Mixed-type variables} combining natural language, discrete choices, and continuous parameters require specialized operators that evolutionary algorithms naturally provide. The \textbf{non-differentiable objective} function involves executing complex agent pipelines without gradient information, while evolution requires only fitness evaluations. The \textbf{multimodal landscape} contains multiple local optima corresponding to different prompting strategies, making population-based search essential for simultaneously exploring multiple regions. \textbf{Expensive evaluation} requires complete benchmark execution, but evolutionary algorithms exploit population information efficiently through hierarchical evaluation strategies. Finally, \textbf{semantic constraints} mean not all configurations are valid; LLM-powered operators maintain validity while exploring variations, unlike random perturbations that often produce meaningless results.

These properties, combined with the success of evolutionary methods in neural architecture search and AutoML, motivate our approach. The semantic awareness of LLM-powered genetic operators enables meaningful exploration while respecting natural language constraints.
\section{The Artemis Platform}\label{sec:artemis}

{This work leverages Artemis, an AI-powered code intelligence platform developed by TurinTech AI. Its foundations trace to earlier research on genetic algorithms for code generation~\cite{artemis_original}, initiated before the emergence of modern LLMs. Since then, it has evolved into a production system powered by TurinTech's Intelligence Engine, a sophisticated orchestration layer that coordinates multi-agent workflows, evolutionary search strategies, and machine learning models to optimize and validate AI-generated code. Artemis has demonstrated strong capability across diverse software engineering challenges, notably achieving a 57\% resolution rate on SWE-Bench~\cite {swe-bench}. It has been applied to real-world code intelligence tasks~\cite{Artemis-use-cases}.

While genetic algorithms have long been applied to optimization problems, their effective use in LLM-based agent systems poses distinct challenges. Critical decisions include determining when to invoke expensive benchmark evaluations versus lightweight proxy metrics, calibrating confidence signals from heterogeneous LLM evaluators, and designing semantic mutation and crossover operators that balance exploration with component validity. The platform's orchestration mechanisms address these challenges using heuristics refined through extensive optimization runs.
In addition to these algorithmic elements, Artemis offers a no-code interface that enables practitioners to apply evolutionary optimization to agent systems without requiring expertise in evolutionary computation. The platform workflow consists of three key stages:}

\begin{figure}[H] 
	\includegraphics[width=\linewidth]{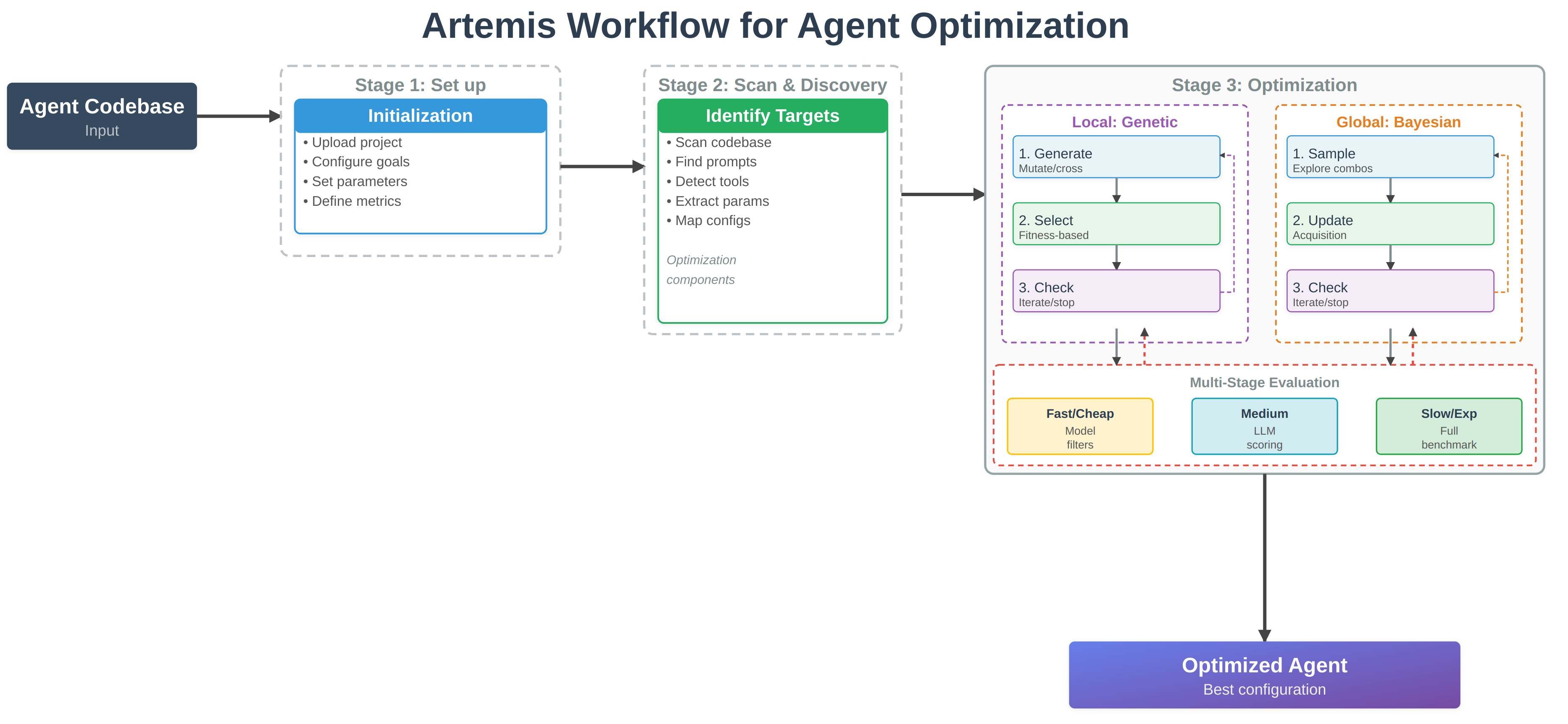}
\caption{Artemis interface showing automatic
	component discovery based on natural-language optimization goals, eliminating manual file specification.} 
\label{fig:platform_interface}
\end{figure}

\textbf{1. Project Setup.} {Users upload their codebase, define optimization objectives in natural language (e.g., ''maximize accuracy while reducing API calls''), specify the evaluation benchmark, and configure search parameters (including choosing the LLMs to invoke).}

\textbf{2. Component Discovery.}
{Artemis automatically analyzes the codebase structure to identify optimizable components such as prompts, tool descriptions, model parameters, and execution settings. The platform supports both \emph{global criteria} (e.g., find all prompts,'' locate all tool descriptions'') and \emph{natural language queries} (e.g., ''find components related to error handling''), using semantic search to locate relevant code without manual specification.}

\textbf{3. Optimization Strategies.} Once components and objectives are established, Artemis provides two complementary optimization approaches:

\begin{itemize}[itemsep=2pt, topsep=0pt] \item \textbf{Local Optimization:}
			Evolves individual components independently using
			genetic algorithms (GA). Each component undergoes semantic
			mutations and crossovers while maintaining contextual
			validity. It is most suitable for components without strong
			interdependencies.

\item \textbf{Global Optimization:} Uses Bayesian optimization to find optimal
combinations when components interact. Explores the combinatorial space of
component versions to identify synergistic configurations. Essential when
prompt-tool interactions affect performance.  \end{itemize}

\begin{figure}[H] \centering
	\includegraphics[width=.9\linewidth]{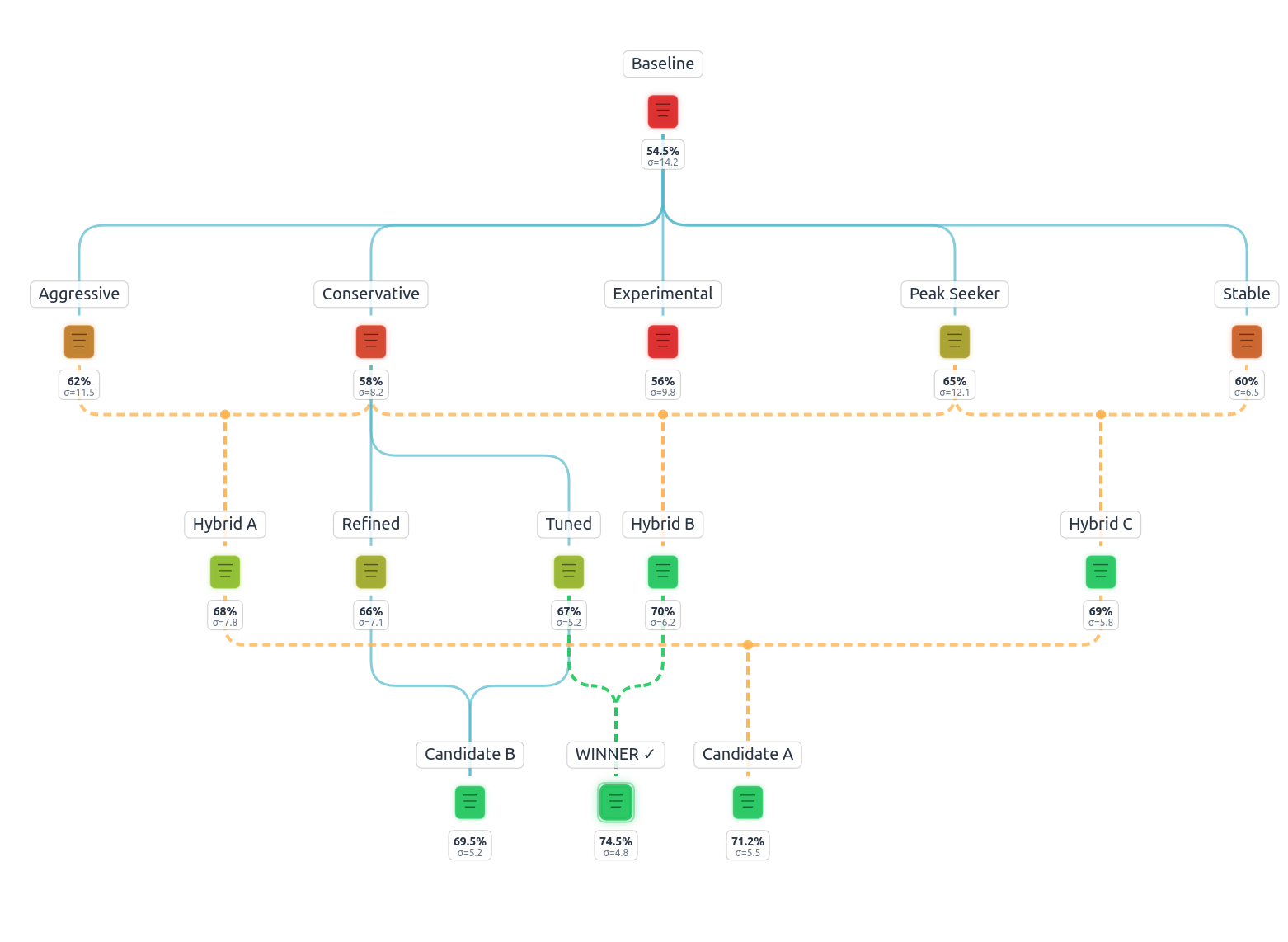}
	\caption{The Artemis semantic GA creates a dynamically evolving search
	tree. LLM ensembles perform mutations and crossovers on components,
	with hierarchical evaluation using cheap filters (LLM scoring) before
	expensive validation (benchmark execution).}
\label{fig:semantic_ga_tree} \end{figure}

The optimization engine employs \textbf{semantic genetic algorithms} where LLM
ensembles perform intelligent mutations that preserve meaning while exploring
variations. Unlike traditional GAs operating on bit strings, Artemis maintains
semantic validity throughout evolution. The \textbf{hierarchical evaluation}
strategy balances efficiency with accuracy: cheap evaluators (LLM-based
scoring, static analysis) filter candidates quickly, while expensive evaluators
(full benchmark runs) validate only promising configurations.

Key advantages include: no coding required (natural language interface),
automatic component discovery (semantic search eliminates manual
specification), intelligent evolution (LLM-powered operators maintain
validity), and black-box optimization (works with any agent architecture
without modifications). By abstracting complex evolutionary algorithms behind an intuitive interface, Artemis makes sophisticated optimization accessible to
practitioners without specialized expertise.

\textbf{Illustration of Workflow.} To illustrate \Artemis's 
distinctive approach, consider optimizing a code performance agent. First upon uploading the codebase, we can identify the optimisation targets based either on tools or an agent using natural language queries:
``I would like to optimise the performance of this codebase, help me find the most important files that I need to change to achieve that.''
\begin{figure}[H] 
	\includegraphics[width=\linewidth]{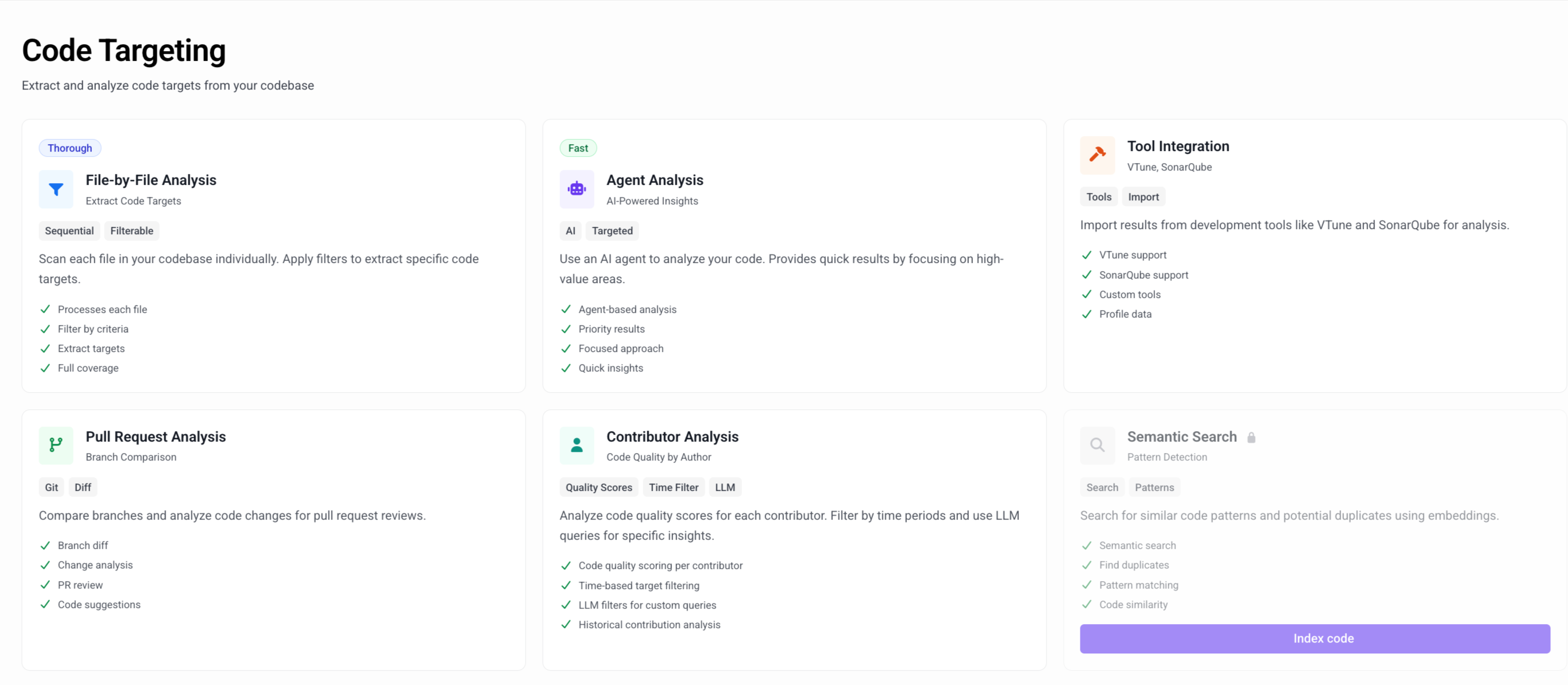}
\caption{Artemis target identification.} 
\label{fig:platform_targets}
\end{figure}

The baseline 
agent uses a generic prompt: ``Refactor this code for better performance based on the following metrics: accuracy, cost, etc.'' 
\begin{figure}[H] 
	\includegraphics[width=\linewidth]{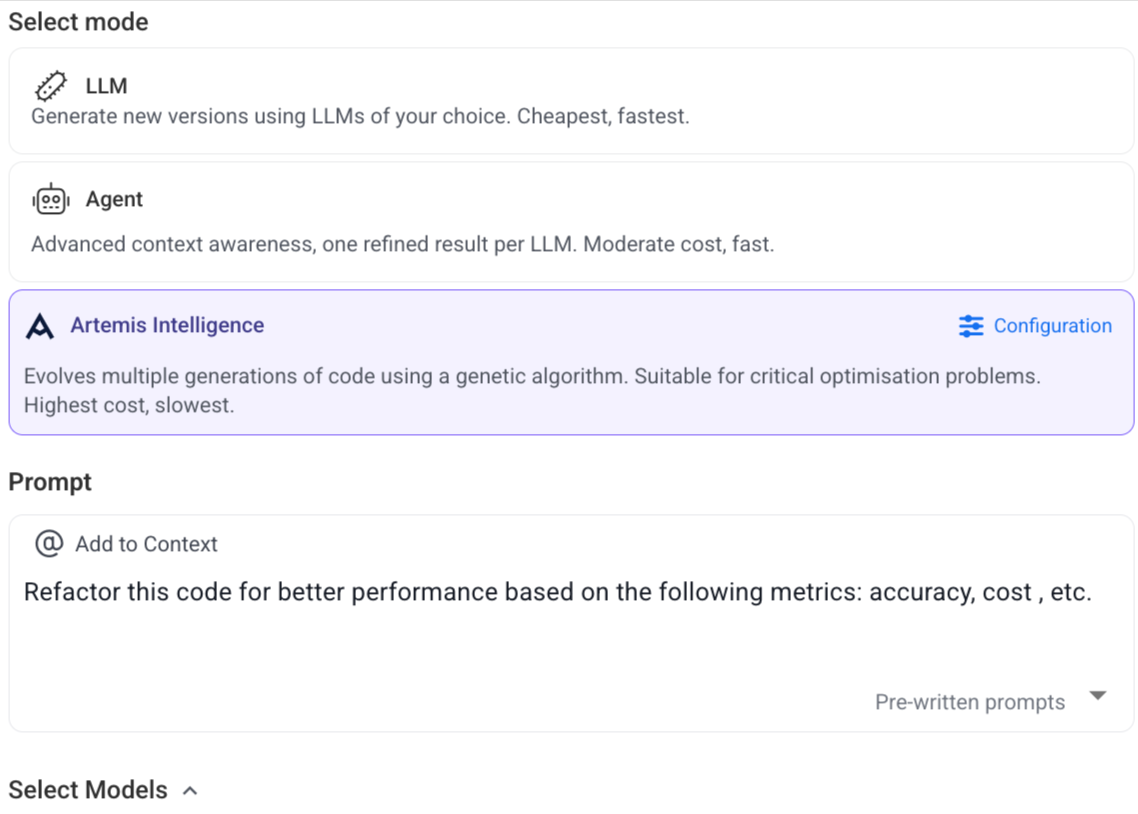}
\caption{Artemis platform optimization prompt.} 
\label{fig:platform_optprompt}
\end{figure}

\Artemis then initializes a population of candidate configurations. It applies semantic genetic operators: LLM-based mutations that generate new versions of code components and crossovers that merge successful elements from different candidates to produce improved variants. Each generation is evaluated on benchmark tasks using a multi-stage evaluation strategy that progresses from cheap, fast techniques (model-based scorers, LLM judges) to expensive, slow measurements (actual benchmark execution on real tasks), with fitness derived from performance metrics and execution logs. This optimization requires no architectural changes to the agent; only a benchmark script and natural-language optimization goals.

\section{Experimental Setup}\label{sec:setup}

We evaluate Artemis across four diverse agent systems spanning competitive
programming, code optimization, and mathematical reasoning. Each
agent-benchmark pair tests different capabilities of the optimization framework,
ranging from complex algorithmic problem-solving to performance tuning and logical
reasoning.

\subsection{ALE Agent on AtCoder Heuristic Contest}

{
This use case addresses competitive programming tasks using the \textbf{ALE Benchmark}, which consists of 40 challenging heuristic-optimization problems from AtCoder's Heuristic Contest. The agent tackles algorithmic problems requiring sophisticated reasoning and solution generation through a multi-stage pipeline: problem analysis, solution planning, code generation, and iterative refinement based on test results.

The primary evaluation metric is acceptance rate, calculated as the fraction of problems passing all test cases (fully accepted problems / total problems), ranging 0.0--1.0. The benchmark also computes average performance (mean competitive programming rating from AtCoder's formula, converting each problem's rank to a performance score ranging 200--3200) and average rank (mean ranking position vs. historical submissions). However, optimization was performed exclusively for acceptance rate; performance and rank metrics were automatically calculated by the benchmark but never used in optimization logic, improving only indirectly through better algorithmic choices for correctness.

We optimize two components: system prompts and search strategies. For prompt optimization, we employ evolutionary prompt engineering to evolve prompts from simple Chain-of-Thought (CoT) approaches to include sophisticated self-correction checklists. For search optimization, we apply genetic algorithms with advanced search strategies (such as beam search and taboo memory) for systematic exploration of the solution space.
}

\vspace{0.3em} \noindent\rule{\textwidth}{0.4pt} \vspace{-0.5em}
\noindent\begin{tabular}{@{}p{2.5cm}p{10.5cm}} \textit{Task:} & Competitive
	programming with 40 algorithmic problems\\ \textit{Components:} &
	System prompts and search strategies\\ 
    \textit{Metric:} & Acceptance
rate\\ \textit{Cost:} & \$24-26 per evaluation run \end{tabular}

\vspace{0.3em} \noindent\rule{\textwidth}{0.4pt} \vspace{-0.5em}

\subsection{Mini-SWE Agent on SWE-Perf}

{This use-case addresses the unique challenge of \textbf{software performance optimization}, which demands a deep understanding of algorithmic complexity, data structures, and system bottlenecks to devise creative solutions that yield measurable speedups. The central difficulty lies in the vast, nuanced search space of possible optimizations (algorithmic, memory, parallelization, etc.), where one needs to tailor effective strategies to specific codebases. To tackle this, we focus on the \textbf{SWE-Perf benchmark}, which presents 140 performance optimization instances across nine major Python projects (e.g., \texttt{scikit-learn}, \texttt{requests}), challenging agents to generate correct, speed-improving code patches. We utilized the \textbf{mini-SWE-agent} framework, a lightweight, single-LLM agent for autonomous coding tasks. We demonstrate how \textbf{Artemis Intelligence} can systematically employ prompt engineering, specifically using genetic algorithms, to evolve highly effective optimization strategies and unlock significant performance gains on the benchmark.}

\vspace{0.3em} \noindent\rule{\textwidth}{0.4pt} \vspace{-0.5em}

\noindent
\begin{tabular}{@{}p{2.5cm}p{10.5cm}} \textit{Task:} & Python code
	optimization across 140 functions from 9 popular projects\\
	\textit{Components:} & YAML configuration templates and optimization
	instructions\\ 
    \textit{Metric:} &
	Performance score improvement\\ \textit{Projects:} & astropy,
	matplotlib, seaborn, requests, xarray, pylint, scikit-learn, sphinx,
	sympy\\ \textit{Base LLM:} & {Claude 3.5 Sonnet}\\
\textit{Cost:} &  {\$30-60 and 20-30 hours evaluation time for a
full benchmarking}\\ \end{tabular}

\vspace{0.3em} \noindent\rule{\textwidth}{0.4pt} \vspace{-0.5em}

\subsection{\large CrewAI Agent on Math Odyssey}

{
The use case centers on solving the highly challenging \textbf{Math Odyssey benchmark}, a collection of 387 diverse mathematics problems spanning algebra, geometry, and number theory. The problems range in difficulty from simple multiple-choice questions to complex word problems, serving as proxies for real-world tasks demanding complex text comprehension and precise calculations. The agent processes natural language problem statements and generates step-by-step mathematical solutions, final calculations, and answers through a streamlined two YAML file system without external tools.

Our core challenge is optimizing \textbf{CrewAI}, a capable general-purpose agent framework with over $10^{20}$ possible configurations due to its numerous prompts and parameters. The components optimized include YAML configuration files containing 12 prompts and 20+ parameters.

We evaluate two primary metrics: accuracy and cost (tokens). Accuracy is measured by counting the number of correct problems per evaluation run (30 or 50 problems), where the agent's calculated answer is compared against an answer key. Token count tracks total tokens (input + output) across two API requests per problem; one for the researcher agent solving the problem and one for the reporting analyst agent evaluating correctness. The experimental setup includes control groups and an optimized test group evaluated using two configurations (30×10 and 50×6 problems) to test whether stratified sampling with a sample size of 30 adequately represents the true distribution of the entire 387-problem corpus.
}

\subsection*{Summary}

\vspace{0.3em} \noindent\rule{\textwidth}{0.4pt} \vspace{-0.5em}
\noindent
\begin{tabular}{@{}p{2.5cm}p{9.51cm}} 
\textit{Task:} & Natural
	language processing and mathematical reasoning across 387 problems.\\
\textit{Components:} & YAML configuration files containing 12 prompts,
	and 20+ parameters\\ 
\textit{Metric:} & Accuracy, Cost (tokens)\\ 
\textit{Evaluation:} & stratified sample of 30 problem capabilities
\end{tabular}

\subsection{MathTales-Teacher Agent on Math Problem Solving}
{
The MathTales-Teacher agent is designed to solve primary-level mathematics problems, and its generated solution steps and final answers are used to create story-based mathematics learning materials for primary-level children \cite{zhang2025multimodal}. The agent is powered by the Qwen2.5-7B model and follows a Reason-and-Act (ReAct) architecture. However, the agent driven by a small parameter language model (SLM) faces two challenges. First, the agent may become stuck during execution and fall into a loop, typically because it fails to select the correct action based on contextual information. Second, the model may produce confident but incorrect numerical calculations. To address these challenges, we apply Artemis to automatically optimise prompt configuration within the current agent's action logic. This configuration file contains the prompts for each action the agent can take under the ReAct workflow. For optimisation and evaluation, we use the Grade School Math 8K (GSM8K) benchmark published by OpenAI \cite{cobbe2021gsm8k}, and measure the agent's task completion rate and answer accuracy as the primary metrics. During the optimisation process, we extract a subset of 50 problems from the benchmark as the validation set, select the configuration achieving the best metric performance, and then evaluate this selected version against the baseline on a larger set of 300 problems to measure the performance and generalisation of the optimisation made by Artemis.

\vspace{0.3em} \noindent\rule{0.9\textwidth}{0.4pt} \vspace{-0.5em}
\noindent

\begin{tabular}{@{}p{2.3cm}p{11cm}} 
\textit{Task:} & Primary-level mathematics problem solving.\\
\textit{Components:} & Prompt configuration file containing five prompts. \\ 
\textit{Optimization:} & Genetic algorithm with two generations and a population size of 3. \\ 
\textit{Metric:} & Accuracy, Completeness. \\ 
\textit{Validation:} & Stratified sample of 50 problems from GSM8K. \\
\textit{Evaluation:} & Full benchmark assessment on 300 problems from GSM8K.
\end{tabular}

\vspace{0.3em} \noindent\rule{0.9\textwidth}{0.4pt}
}

Below is the summary of the agent characteristics:

\vspace{0.3em} \noindent\rule{\textwidth}{0.4pt}

\begin{table}[H] \centering \caption{Summary of agent characteristics and
	optimization parameters} \label{tab:agent_summary} \small
	\begin{tabular}{lccccc} \toprule \textbf{Agent} & \textbf{Full Benchmark} &\textbf{Test Cases} &
		\textbf{Primary Metric}
        \\ \midrule ALE & 40 problems &	- &	Acceptance rate 
        \\ Mini-SWE & 140 functions & - & Performance score 
		\\ CrewAI & 387 problems & 15 problems & Accuracy (\%), Cost(tokens) 
        \\ MathTales & 300 problems & 50 problems & Accuracy and Completeness 
		\\ \bottomrule \end{tabular}
\end{table}
\section{Results}\label{sec:results}

We present detailed results for each agent-benchmark pair, followed by
aggregate performance analysis across all systems. Our evaluation reveals both
the strengths and limitations of automated optimization through Artemis.

\subsection{ALE Agent: Competitive Programming}

The ALE Agent tackles competitive programming problems from AtCoder's Heuristic
Contest, requiring sophisticated algorithmic reasoning and solution generation.
The agent employs a multi-stage pipeline: problem analysis, solution planning,
code generation, and iterative refinement based on test results. The plots normalize performance and rank metrics while acceptance rate remains on a 0--1 scale. Average performance values (raw range ~200--3200) are normalized by dividing by the maximum (~3200), while average rank values are normalized by total participants per problem (typically 3700--5000). These normalized metrics provide relative comparisons but were not used in optimization, which focused exclusively on acceptance rate.

\begin{figure}[!ht]
    \centering
    \begin{minipage}[t]{0.48\textwidth}
        \centering
        \colorbox{red!45}{\parbox{0.95\textwidth}{\centering\textbf{Before}}}
        \vspace{0.2cm}
        \begin{quote}
            \textit{``Generate a solution for the given problem. Consider edge cases and optimize for performance. Implement the algorithm efficiently.''}
        \end{quote}
    \end{minipage}%
    \hfill
    \begin{minipage}[t]{0.48\textwidth}
        \centering
        \colorbox{green!45}{\parbox{0.95\textwidth}{\centering\textbf{After}}}
        \vspace{0.2cm}
        \begin{quote}
            \textit{``Decompose the problem into subcomponents: (1) identify input/output patterns, (2) detect algorithmic category (graph, DP, greedy), (3) enumerate edge cases explicitly (n=0, n=1, maximum bounds), (4) implement with clear variable naming and modular functions. Validate against sample inputs before submission.''}
        \end{quote}
    \end{minipage}
    \caption{Comparison of original (left) and Artemis-optimized (right) configuration examples.}
    \label{fig:config_comparison}
\end{figure}

\begin{figure}[!ht] \centering
	\includegraphics[width=\textwidth]{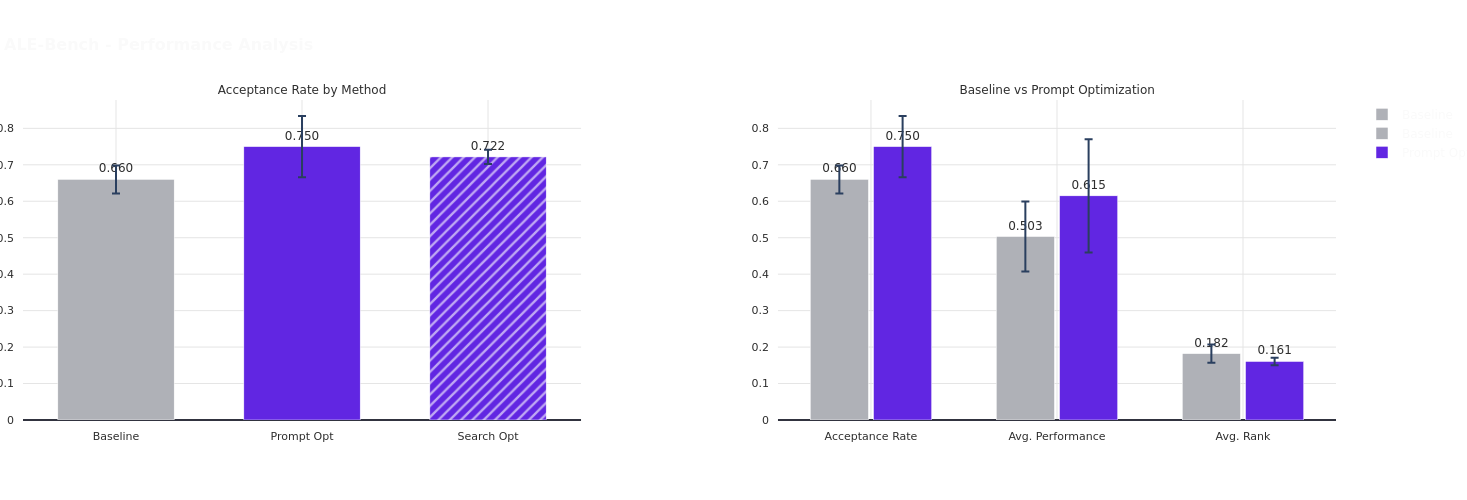}
	\caption{ALE Agent performance on AtCoder Heuristic Contest. Left:
	Comparison of optimization methods showing acceptance rates for
	baseline (0.660), prompt optimization (0.750), and search optimization
	(0.722). Right: Multi-metric performance comparison across acceptance
	rate, average performance, and ranking metrics.}
\label{fig:ale_agent_analysis}
\end{figure}

{We evaluated two optimization strategies: prompt optimization, evolving prompts from simple Chain-of-Thought approaches to include sophisticated self-correction checklists, and search optimization using genetic algorithms with advanced search strategies (beam search and taboo memory). The ALE Agent achieved a \textbf{$13.6\%$ improvement} in
acceptance rate through prompt optimization, rising from $66.0\%$ (95\% CI:
[0.594, 0.726]) to $75.0\%$ (95\% CI: [0.689, 0.811]). The search-based
optimization strategy yielded a $9.3\%$ improvement, reaching $72.2\%$ (95\%
CI: [0.661, 0.783]). While neither reached statistical significance at $\alpha
= 0.05$ ($p = 0.10$ for both), the practical improvements are substantial for
competitive programming contexts where each percentage point of acceptance rate
translates to additional solved problems and higher rankings. The optimization
process required 411.2 hours for prompt optimization and 260.5 hours for search
optimization, with costs of $\$24-26$ per evaluation run.}

{Key improvements focused on structured problem decomposition
and explicit edge case handling. The optimized prompts guide the agent through systematic analysis phases rather than attempting immediate solution
generation, resulting in more robust and correct implementations.}

\subsection{Mini-SWE Agent: Code Performance Optimization}

Code performance optimization differs fundamentally from code testing or bug fixing, requiring a deep understanding of algorithmic complexity,
data structures, and system bottlenecks while maintaining correctness
\cite{gong2025language}. This difference makes it an ideal testbed for Artemis: the vast
search space of optimization strategies (algorithmic improvements, memory
efficiency, parallelization, caching) and the need for codebase-specific
solutions create a complex optimization landscape where evolutionary prompt
engineering can discover non-obvious approaches.

We evaluate on SWE-Perf \cite{he2025swe}, which tests agents on
140 optimization instances across nine major Python projects (astropy, requests,
scikit-learn, etc.). The Artemis-optimized configuration emphasizes targeting
the single most critical performance bottleneck, systematic complexity analysis
before optimization, data structure selection based on access patterns, and
domain-specific techniques like vectorization and caching.

\noindent
\subsection*{YAML Configuration Snippet:} 

\begin{figure}[!ht] 
    \centering
    \begin{minipage}[t]{0.48\textwidth}
        \centering
        \colorbox{red!45}{\parbox{0.95\textwidth}{\centering\textbf{Before}}}
        \begin{minted}[breaklines, breakanywhere, fontsize=\small, frame=leftline, escapeinside=||]{yaml}
optimization_strategy: 
    focus: "general improvements" 
    methods: 
        - "algorithm optimization"
        - "caching"
        \end{minted}
    \end{minipage}
    \hfill
    \begin{minipage}[t]{0.48\textwidth}
        \centering
        \colorbox{green!45}{\parbox{0.95\textwidth}{\centering\textbf{After}}}
        \begin{minted}[breaklines, breakanywhere, fontsize=\small, frame=leftline, escapeinside=||]{yaml}
optimization_strategy: 
    focus: bottleneck-driven optimization
    methods: 
        - complexity analysis before optimization
        - data structure selection based on access patterns
        - vectorization for numerical operations
        - lazy evaluation where applicable
    validation: ensure identical output with performance benchmarking
\end{minted}
\end{minipage}
    \caption{Comparison of original (left) and Artemis-optimized (right) YAML configuration strategies.}
    \label{fig:yaml_comparison}
\end{figure}

\begin{figure}[h]
      \centering
	\includegraphics[width=\textwidth]{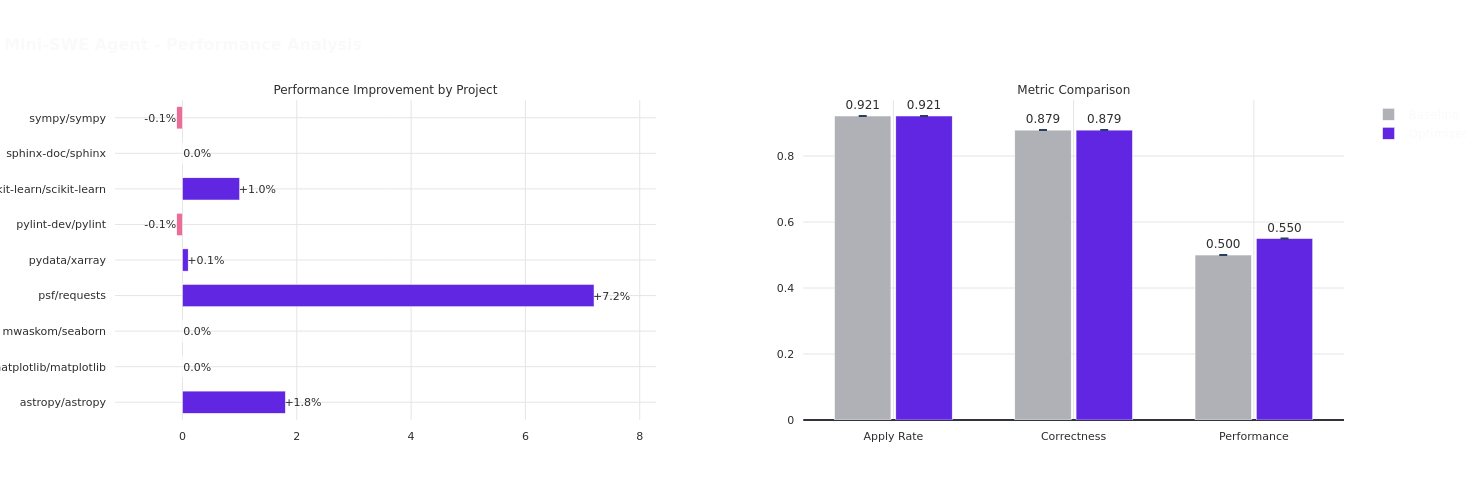}
	\caption{Mini-SWE Agent performance on SWE-Perf benchmark. Left:
	Project-specific improvements in percentage points, showing the largest gains for requests
	($+7.2$ points, $36.1\%\rightarrow43.3\%$) and astropy ($+1.8$ points). Right: Overall metrics comparison 
    showing consistent apply rate and correctness with improved performance
	score.} \label{fig:swe_agent_analysis} 
      \label{figurelabel}
\end{figure}

{The Mini-SWE Agent demonstrated a statistically significant
\textbf{10.1\% performance improvement} (p < 0.05) using the Mann-Whitney U test,
with apply rate and correctness maintained at 92.1\% and 87.9\%. Project-level
results varied: requests showed +20\% relative improvement
(36.1\%$\rightarrow$43.3\%), scikit-learn +29\% (3.5\%$\rightarrow$4.5\%), and
astropy +62\% (2.9\%$\rightarrow$4.7\%). These project-specific gains
demonstrate the agent's ability to identify optimization opportunities in
production libraries.}

\subsubsection*{Case Study: \textit{astropy} Array Comparison.}

{We examine the optimization of \texttt{report\_diff\_values()}
in \texttt{astropy/utils/diff.py}, demonstrating the agent's systematic
approach. The first example in the figure~\ref{fig:astropy_comparison} shows the original implementation
with repeated string operations and generic array handling.
The second Example~\ref{fig:astropy_comparison} presents the optimized version with six key
improvements: (1) early identity check, (2) cached string operations via
\texttt{\_get\_indent()}, (3) optimized array dtypes (float64/int64), (4) batch
processing for large arrays, (5) vectorized type-specific comparisons via
\texttt{\_fast\_array\_compare()}, and (6) contiguous memory layout for cache
efficiency.}

{The optimization completed in 9 hours across three generations with
a population size of 3. The improved configuration emphasizes systematic bottleneck
analysis before applying optimizations, leading to more targeted and effective
performance improvements.}

\subsection{CrewAI Agent: Mathematical Reasoning}

The CrewAI Agent handles conversational mathematical reasoning tasks through a streamlined two YAML file-system without external tools. The agent processes natural language problem statements and generates step-by-step mathematical solutions, final calculations, and answers. Token counts represent total tokens (input + output) across two API requests per problem: one for the researcher agent solving the problem and one for the reporting analyst agent evaluating correctness. The reduction in total tokens primarily reflects decreased output token generation, directly translating to cost savings. Accuracy is the fraction of problems answered correctly per run. The baseline configuration was evaluated over 10 runs of 30 problems (30×10). The optimized configuration was evaluated in two designs, 30×10 and 6 runs of 50 problems (50×6), to validate that stratified sampling with 30 problems adequately represents the full 387-problem corpus distribution. We report the 30×10 evaluation as the primary result, since it matches the baseline design, and report the 50×6 evaluation alongside it.

Below is an example of a very simplistic original prompt and \Artemis-optimization attempt:

The example below shows the result of this adjustment on the code:

\begin{figure}[!ht] 
\centering
\includegraphics[width=0.9\textwidth]{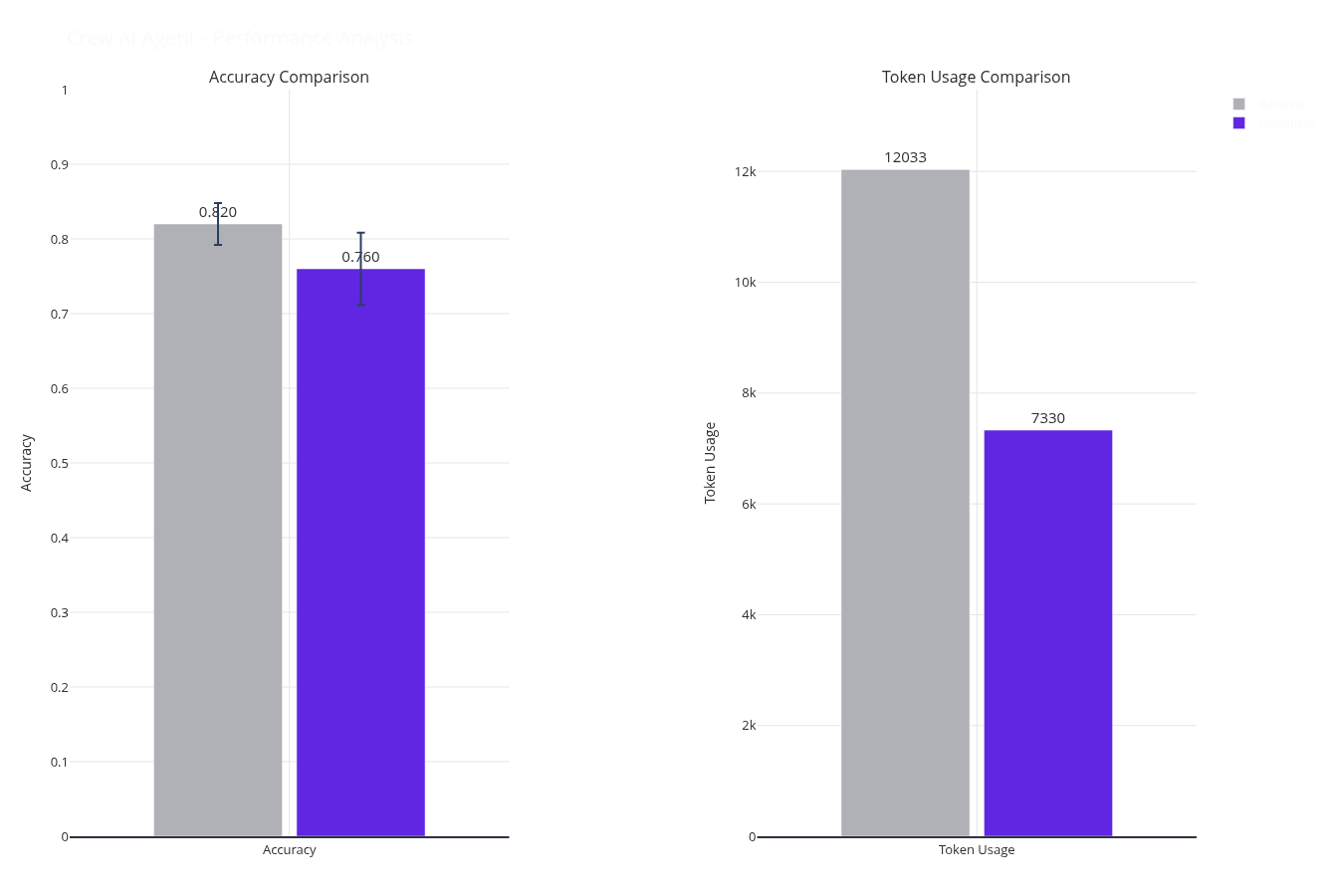}
\caption{CrewAI Agent on Math Odyssey benchmark. Left: Average accuracy of the original and \Artemis-optimized configurations, each over 10 runs $\times$ 30 problems; error bars show 95\% confidence intervals. Right: Average total tokens per problem.}
\label{fig:math_agent_analysis}
\end{figure}

The CrewAI Agent showed a significant \textbf{39.1\%} \textbf{decrease} in the average number of tokens used per problem (12033 $\rightarrow$ 7330, p $< 10^{-6}$), with a corresponding \textbf{33.2\% decrease} in the median cost per problem. This came with a small but statistically significant \textbf{7.3\% decrease} in the average rate of correct problems (0.82 $\rightarrow$ 0.76, p $= 0.027$). The 50×6 evaluation of the same optimized configuration gives a consistent picture: a 36.9\% reduction in mean tokens (7590), a 36.2\% reduction in median tokens, and an accuracy of 0.78 (-4.9\%, p $= 0.277$), a decrease that is not statistically significant at that sample size. The original configuration was already well-tuned for performance regarding correctness, with little room for improvement, so we focused on optimizing for cost. We find that Artemis can identify significant cost-optimization opportunities in tasks where there is room for cost efficiency. In this case, Artemis modified each prompt and set reasonable token limits. The result is that problems with exceptional expense are intentionally failed at zero expense. In contrast, medium-difficulty problems are executed more efficiently, as evidenced by the changes in average and median prices. Correspondingly, it appears that the exceptionally expensive problems are also the problems that were likely to be incorrect, given that the decrease in performance is small relative to the reduction in cost.

\subsection{MathTales-Teacher Agent: Primary-level Math Problem Solving}
MathTales-Teacher agent is designed to solve primary-level mathematics questions and provide a foundation for generating mathematical stories. The agent is powered by Qwen2.5-7B in our experiments and processes natural-language mathematics problems, producing step-by-step solutions and final answers. Figure \ref{fig:mathtale_prompt_comparison} shows the prompt optimisation performed by Artemis for solving action in the agent.

\begin{figure}
    \centering
    \includegraphics[width=0.99\linewidth]{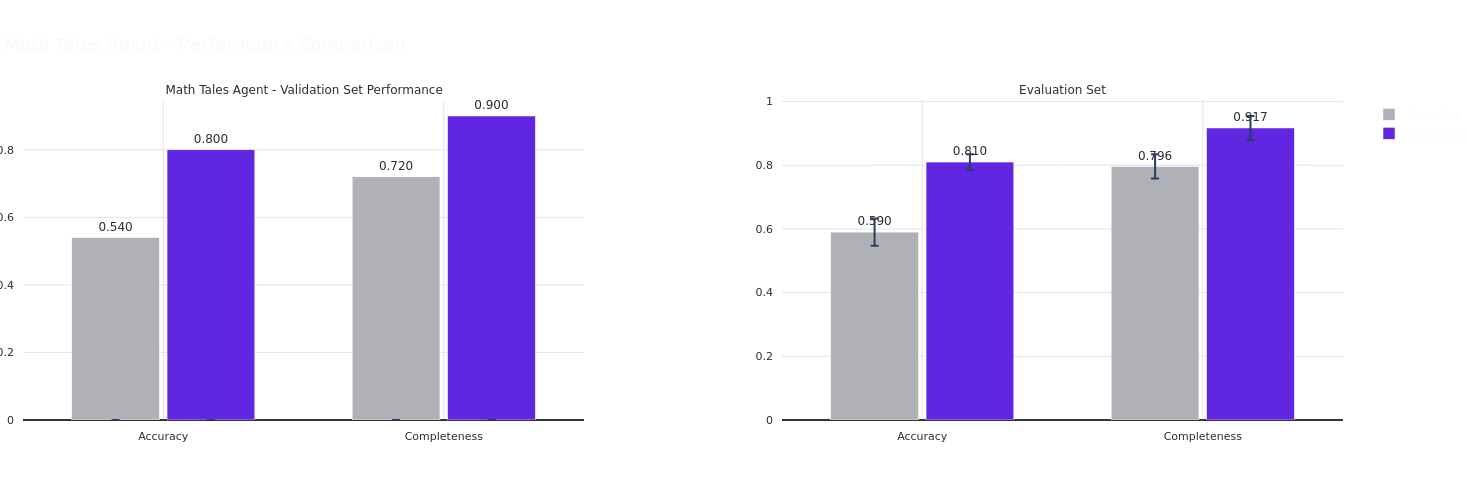}
    \caption{MathTales-Teacher agent on GSM8K benchmark. Left: Comparison of Completeness and Accuracy on the validation set. Right: Average Accuracy and Completeness comparison across 3 evaluation runs.}
    \label{fig:mathtale_agent_analysis}
\end{figure}

One can execute the \textbf{MathTales} locally and does not rely on any commercial models; hence, we do not consider token usage or cost in our experiments. Instead, we focus on evaluating the problem completion rate and answer accuracy. Figure \ref{fig:mathtale_agent_analysis} presents the performance of the original prompt configuration (baseline) and the Artemis-optimised configuration on both the validation set (50 mathematics problems) and the evaluation set (300 mathematics problems). On the validation set, the optimised agent improves the question completion rate from 0.72 to 0.90 (+25.0\%) and the answer accuracy from 0.54 to 0.80 (+48.1\%). Across three repeated runs on the evaluation set, the optimised agent achieves an average completion rate of 0.917 versus 0.796 for the baseline (+15.2\%) and an average accuracy of 0.81 versus 0.59 (+37.3\%). As elsewhere in this paper, improvements are relative changes; in absolute terms the evaluation-set gains are 12.1 and 22 percentage points, respectively. These findings demonstrate that the optimisation performed by Artemis generalises effectively to broader problem distributions within the GSM8K benchmark.

\subsection{Aggregate Performance Analysis}

\begin{figure}[!ht] 
\centering
\includegraphics[width=\textwidth]{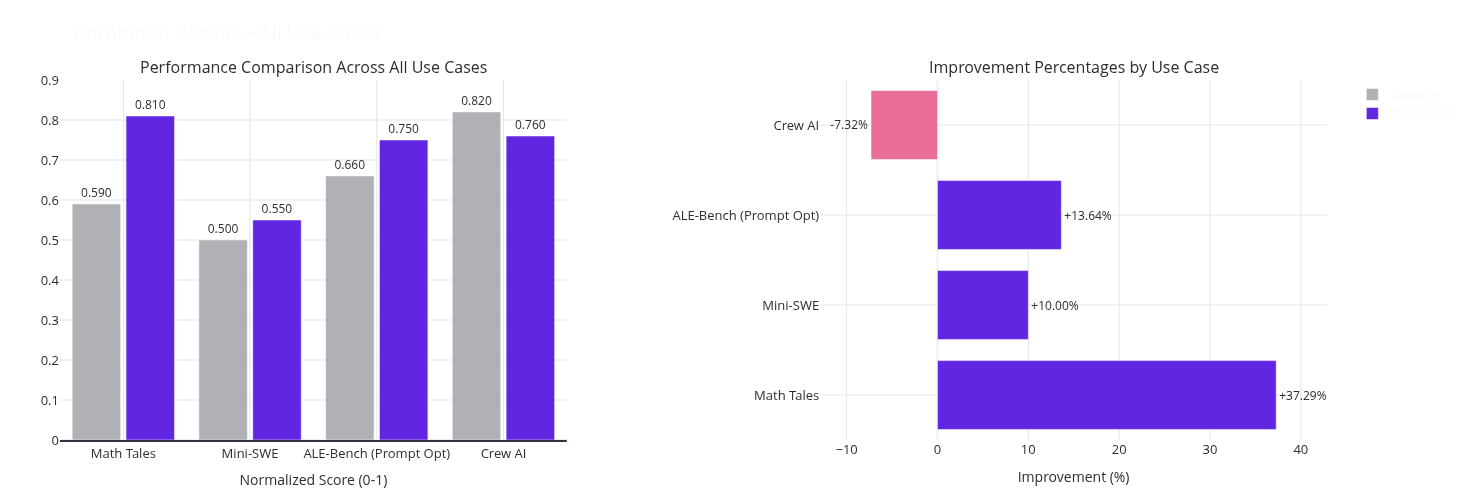}
\caption{Global performance comparison across all four agents.}
\label{fig:global_performance_comparison}
\end{figure}

\begin{table}[!ht] 
\centering
\caption{Summary of optimization results across all evaluated agents}
\label{tab:optimization_summary}
\begin{tabular}{lcccc}
\toprule
\textbf{Agent} & \textbf{Baseline} & \textbf{Optimized} & \textbf{Improvement} & \textbf{p-value} \\
\midrule
ALE (Prompt) & 0.660 & \textbf{0.750} & +13.6\% & 0.10 \\
ALE (Search) & - & \textbf{0.722} & +9.3\% & 0.10 \\
Mini-SWE & 0.891 & \textbf{0.981} & +10.1\% & <0.005 \\
CrewAI (Accuracy, 30×10) & \textbf{0.82} & 0.76 & -7.3\% & 0.027 \\
CrewAI (Accuracy, 50×6) & \textbf{0.82} & 0.78 & -4.9\% & 0.277 \\
CrewAI (Token Cost, 30×10) & 12033 & \textbf{7330} & 39.1\% & <10$^{-6}$ \\
CrewAI (Token Cost, 50×6) & 12033 & \textbf{7590} & 36.9\% & <10$^{-6}$ \\
MathTales (Accuracy) & 0.59 & \textbf{0.81} & +37.3\% & <0.001 \\
MathTales (Completeness) & 0.796 & \textbf{0.917} & +15.2\% & <0.001 \\
\bottomrule
\end{tabular}
\end{table}

Our comprehensive evaluation across four diverse agent systems demonstrates Artemis's effectiveness in automated optimization, though with notable variations in success patterns and optimization objectives. Three out of four agents showed statistically significant improvements in their primary performance metrics, with improvements ranging from 9.3\% to 39.1\% for successful optimizations.

The ALE Agent achieved a 13.6\% relative improvement in acceptance rate (0.66 → 0.75 through prompt optimization), demonstrating that evolutionary prompt engineering can substantially enhance competitive programming performance even when starting from a moderate baseline (66\%). The Mini-SWE Agent showed statistically significant improvement (p < 0.005) with a 10.1\% gain in performance score, validating Artemis's ability to optimize code performance tasks with well-defined metrics. The MathTales-Teacher Agent exhibited strong improvements in both accuracy (0.59 → 0.81, +37.3\%, p < 0.001) and completion rate (0.796 → 0.917, +15.2\%, p < 0.001), indicating that Artemis effectively addresses challenges specific to small language models, such as execution loops and calculation errors.

The CrewAI Agent presents an interesting case study in multi-objective optimization. While accuracy showed a small but statistically significant decrease (-7.3\%, p = 0.027 in the 30×10 evaluation; -4.9\%, p = 0.277 in the 50×6 evaluation), the agent achieved a dramatic 39.1\% reduction in token cost (p < 10$^{-6}$), demonstrating Artemis's capability to optimize for cost efficiency when performance is already well-tuned. This trade-off reflects the optimization objective: the original configuration prioritized correctness, leaving substantial room for cost reduction through prompt refinement and token limit adjustments.

Statistical significance patterns reveal important insights: improvements were statistically significant (p < 0.05) for Mini-SWE, CrewAI's cost reduction, and MathTales-Teacher (both accuracy and completeness, p < 0.001), while ALE improvements, though substantial, did not reach significance (p = 0.10) due to the high variance inherent in competitive programming evaluations. The total optimization time varied significantly across agents: 671.7 hours for ALE Agent (due to expensive competitive programming evaluations), 9 hours for Mini-SWE, and variable time for CrewAI and MathTales, reflecting the computational cost differences between benchmark types.

\subsection{Key Insights and Limitations}

Our comprehensive evaluation reveals several critical insights about automated agent optimization:

\textbf{Optimization Effectiveness Factors.} Success depends on three key factors: (1) the quality of the initial configuration; poorly tuned agents show greater improvement potential; (2) the nature of the task; well-defined metrics like acceptance rate or performance score enable better optimization than subjective reasoning tasks; and (3) the optimization strategy; prompt optimization excels for instruction clarity while search strategies work for systematic exploration.

\textbf{Computational Trade-offs.} While optimization requires significant computational resources (hundreds of hours for complex benchmarks), the resulting improvements often justify the investment. Importantly, configurations optimized for performance typically maintain similar runtime cost to baselines while delivering better performance. The hierarchical evaluation strategy in Artemis helps manage these costs by efficiently filtering candidates. Configurations that already exhibit high performance may also be candidates for optimization in other metrics, such as cost. Artemis can help automate this search using local optimizations with a variety of prompt configurations targeted toward specific metrics.

The results validate Artemis as a practical framework for automated agent optimization, particularly for agents with clear performance metrics and room for improvement. The system's ability to discover non-obvious optimizations through semantic mutations demonstrates the value of evolutionary approaches for natural language components. However, the mixed results also highlight that automated optimization is not universally beneficial; practitioners should assess their agents' baseline quality and task characteristics before investing in optimization efforts.
\section{Discussion}\label{sec:discussion}

Our experiments with Artemis demonstrate successful optimization, achieving
$\textcolor{blue!70!black}{13.6\%}$ improvement for the ALE Agent,
$\textcolor{blue!70!black}{10.1\%}$ for the Mini-SWE Agent, and
$\textcolor{blue!70!black}{39.1\%}$ cost reduction for the CrewAI Agent. {The MathTales-Teacher
Agent achieved $\textcolor{blue!70!black}{37.3\%}$ accuracy improvement on
primary-level mathematics problems.} These results provide insights into the
conditions under which evolutionary optimization of LLM agents succeeds.

The most striking pattern is the relationship between initial configuration
quality and optimization success. Both ALE and Mini-SWE started with generic,
underspecified configurations. The ALE Agent's original prompts used vague
instructions like "consider edge cases," which Artemis transformed into
explicit decomposition steps and systematic validation. Similarly, the Mini-SWE
Agent's generic "general improvements" strategy became a targeted
"bottleneck-driven optimization" approach. {The MathTales-Teacher
Agent's simplistic prompts were enhanced with explicit verification steps and
decomposition strategies, addressing challenges in both task completion and
numerical accuracy.} Agents with more refined initial configurations may
require different optimization strategies or longer convergence times to show
improvements.

Task characteristics also influenced outcomes. Competitive programming (ALE)
and code optimization (Mini-SWE) have clear, objective metrics; acceptance rates
and performance scores that provide unambiguous feedback for evolutionary
search. Mathematical reasoning (CrewAI), despite having accuracy and cost metrics, involves more subjective problem-solving approaches where the original human-crafted prompts may already encode effective strategies. This suggests that Artemis works best for tasks with objective solving approaches, well-defined success criteria, and measurable outcomes.

The choice of optimization strategy proved more important than anticipated. For
the ALE Agent, prompt optimization ($\textcolor{blue!70!black}{13.6\%}$)
significantly outperformed search-based optimization
($\textcolor{blue!70!black}{9.3\%}$), despite the search being more systematic.
This indicates that for complex reasoning tasks, improving instruction clarity
yields greater benefits than exploring parameter spaces. The 411.2 hours
invested in prompt optimization, while substantial, produced configurations
that maintain similar per-run costs ($\$25.06$ vs $\$24.59$) while delivering
superior performance.

Our results also reveal important limitations. The Mini-SWE Agent showed high
variance across projects; from $\textcolor{blue!70!black}{+7.2}$ percentage points
(36.1\% $\rightarrow$ 43.3\%) for requests to $\textcolor{blue!70!black}{-0.1}$ percentage points for pylint, indicating that optimizations
may be specific to evaluation datasets rather than universally applicable. This
raises concerns about generalization: improvements on benchmarks may not
translate to real-world usage patterns. Additionally, the ALE Agent's
improvements, while practically significant, did not achieve statistical
significance (p = $\textcolor{blue!70!black}{0.10}$), highlighting the
challenge of rigorous validation given expensive evaluations. The CrewAI
framework was also previously optimized for performance alone, and it did not
make significant improvements. In fact, we observed a small decrease in accuracy (0.82 $\rightarrow$ 0.76,
p = $\textcolor{blue!70!black}{0.027}$), which the optimizer accepted in exchange for the reduction in cost. Already controlling high performance on
Math Odyssey, CrewAI may not have been able to improve via tweaks on parameters
and prompts alone. Indicating that optimizations will have limited success in
metrics that are already close to optimal performance. {The
MathTales-Teacher Agent demonstrated that \Artemis effectively optimizes agents
based on smaller open-source models (Qwen2.5-7B), achieving substantial
improvements ($\textcolor{blue!70!black}{37.3\%}$ accuracy,
$\textcolor{blue!70!black}{15.2\%}$ completion rate) that generalize from
validation to evaluation sets. This validates that evolutionary optimization is
not limited to commercial LLM-based agents and can enhance locally-deployed
systems without API costs.}

The computational costs warrant careful consideration. The ALE Agent required
671.7 total hours across all optimization experiments, representing substantial
resource investment. Organizations must weigh these upfront costs against the
expected deployment lifetime and performance gains. For frequently-used agents
or those in competitive domains, a $\textcolor{blue!70!black}{13.6\%}$
improvement may justify the investment. For occasional-use or already
well-performing agents, manual configuration may remain more cost-effective.

These findings suggest practical guidelines for applying automated
optimization. First, assess baseline configuration quality; agents with vague or
generic prompts are better candidates than carefully tuned systems. Second,
ensure clear performance metrics exist; subjective or multi-faceted success
criteria complicate optimization. Third, automated optimization should be
applied to agents and metrics that are not already performing optimally.
Fourth, for reasoning-heavy tasks, prioritize prompt optimization over
parameter search. Finally, validate improvements on held-out data to ensure
generalization beyond training benchmarks.

{\textbf{Practical Impact.} The framework provides a no-code interface
that enables practitioners to apply evolutionary optimization to agent systems
without requiring expertise in evolutionary algorithms. The automated search
process eliminates manual tuning iterations. Performance improvements achieved
through optimization persist across future deployments of the optimized
configuration.}

{\textbf{Reproducibility.} To support validation of our results, we are going to open source the code for all four case study agents (ALE Agent, Mini-SWE Agent, CrewAI Math Agent, MathTales Agent) as supplementary material. While the complete Artemis platform setup cannot be shared, the provided code allows practitioners to examine our agent implementations and verify the reported performance
improvements.}

The results across our four agents reflect the current state of automated agent
optimization. While Artemis demonstrates that evolutionary approaches can discover non-obvious improvements and create efficiency gains in workflow, it is not a universal solution. Success depends on multiple factors: configuration headroom, task characteristics, metric clarity, and computational budget. As LLM agents become more prevalent, understanding these factors will be crucial for deciding when automated optimization offers value versus when human expertise remains essential.
\section{Conclusion, Limitations, and Future Work}\label{sec:conclusion}

This work demonstrated that \Artemis, a general-purpose evolutionary
optimization platform, can effectively optimize LLM agent configurations
without requiring architectural modifications or manual fitness function
design. Through evaluation on four diverse agents spanning competitive
programming, code optimization, and mathematical reasoning, we demonstrated
substantial improvements where optimization potential existed. The ALE agent
achieved 13.6\% improvement in acceptance rate through prompt optimization,
Mini-SWE showed 10.1\% performance gains on code optimization tasks, and
MathTales-Teacher improved accuracy by 37.3\% on primary-level mathematics
problems. These results validate that automated semantic mutations can
transform vague instructions into structured, effective prompts, often
uncovering non-obvious optimizations. The no-code interface democratizes
sophisticated optimization, enabling practitioners to improve agent performance
without specialized evolutionary-algorithms expertise.

Despite these achievements, we acknowledge several limitations. Optimization
effectiveness varies significantly with initial configuration quality; agents
with well-tuned performance baselines may show limited accuracy improvement
potential, though they can still benefit from optimization for alternative
objectives such as cost reduction (CrewAI achieved 39.1\% token reduction
with a 7.3\% accuracy trade-off). For data science and research tasks,
establishing robust validation setups that avoid overfitting to specific
benchmarks remains challenging, potentially limiting the generalizability of
optimized configurations to real-world deployments. Computational costs remain
substantial (671.7 hours for ALE, though reduced to 9-30 hours for other
agents), requiring careful cost-benefit analysis before deployment.
Additionally, statistical significance was not achieved for all improvements
($p = 0.10$ for ALE), suggesting that larger evaluation sets may be necessary
to conclusively validate optimization gains. While the framework supports
multi-objective optimization, effectively balancing trade-offs between
competing objectives (e.g., accuracy vs. cost) remains challenging, as
demonstrated by CrewAI's substantial cost reduction accompanied by a modest
accuracy decrease.

Future work will focus on three key directions. \textbf{First}, we plan to
leverage \Artemis's planning agent, which combines planning and deeper
reasoning capabilities with genetic algorithms and Bayesian optimization; for
more challenging data science optimization tasks. This more powerful
optimization approach could achieve better and more robust results on complex
tasks where deeper reasoning about configuration spaces is required.
\textbf{Second}, developing predictive metrics to assess optimization potential
before expensive trials would enable practitioners to estimate ROI upfront,
potentially through analysis of prompt specificity, parameter diversity, and
configuration entropy. \textbf{Third}, investigating transfer learning across
related agent domains could enable few-shot optimization where configurations
adapt rapidly to new tasks with minimal evaluation, reducing costs by an order
of magnitude while maintaining optimization quality. As LLM agents become
increasingly prevalent in production systems, automated optimization tools like
Artemis will play a crucial role in bridging the gap between theoretical agent
potential and practical deployed performance.


\section*{ACKNOWLEDGMENT}
This work was supported by EU Horizon 2020 Grant 101008280 (DIOR): \url{https://cordis.europa.eu/project/id/101008280}.

\bibliographystyle{plain}
\bibliography{references}

\appendix
\section{Supplementary Results}

\begin{figure}[H]
    \centering
    \begin{minipage}[t]{0.48\textwidth}
        \centering
        \colorbox{red!45}{\parbox{0.95\textwidth}{\centering\textbf{Before}}}
        \begin{minted}[breaklines, breakanywhere, fontsize=\small, frame=leftline, escapeinside=||]{python}
# Basic array comparison without optimizations
if isinstance(a, np.ndarray) and isinstance(b, np.ndarray):
    if a.shape != b.shape:
        indent_str = _get_indent(indent_width) # |\colorbox{red!30}{Uses cached indent strings}|
        fileobj.write(f"{indent_str}Different array shapes:\n")
        # |\colorbox{red!30}{... recursive calls}|
        return False
    
    if np.issubdtype(a.dtype, np.floating) and np.issubdtype(b.dtype, np.floating):
        diff_indices = np.transpose(where_not_allclose(a, b, rtol=rtol, atol=atol))
    else:
        diff_indices = np.transpose(np.where(a != b))
        \end{minted}
    \end{minipage}%
    \hfill
    \begin{minipage}[t]{0.48\textwidth}
        \centering
        \colorbox{green!45}{\parbox{0.95\textwidth}{\centering\textbf{After}}}
        \begin{minted}[breaklines, breakanywhere, fontsize=\small, frame=leftline, escapeinside=||]{python}
    # |\colorbox{green!30}{Early return for identical objects}|
    if a is b:
        return True
    # |\colorbox{green!30}{Fast path for numpy arrays}|
    if isinstance(a, np.ndarray) and isinstance(b, np.ndarray):
        if a.shape != b.shape:
            indent_str = _get_indent(indent_width)
            # |\colorbox{green!30}{Cached indent strings}|
            fileobj.write(f"{indent_str}Different shapes:\n")
            return False
        
        # |\colorbox{green!30}{Optimize dtypes, batch processing}|
        a_view = _optimize_array_dtype(
            np.ascontiguousarray(a))
        # |\colorbox{green!30}{Converts to float64/int64}|
        b_view = _optimize_array_dtype(
            np.ascontiguousarray(b))
        
        if a_view.size > BATCH_SIZE:
            diff_indices = _batch_compare(
                a_view, b_view, rtol=rtol, atol=atol)
            # |\colorbox{green!30}{Memory-efficient chunking}|
        else:
            diff_indices = _fast_array_compare(
                a_view, b_view, rtol=rtol, atol=atol)
            # |\colorbox{green!30}{Vectorized comparison}|
        \end{minted}
    \end{minipage}
    \caption{Comparison of original (left) and optimized (right) astropy array comparison implementations.}
    \label{fig:astropy_comparison}
\end{figure}

\begin{figure}[H]
    \centering
    \begin{minipage}[t]{0.48\textwidth}
        \centering
        \colorbox{red!45}{\parbox{0.95\textwidth}{\centering\textbf{Before}}}
        \begin{minted}[breaklines, breakanywhere, fontsize=\small, frame=leftline, escapeinside=||]{yaml}
researcher:
  role: >
    Math solver.
  goal: >
    Try to research this problem {topic}.
  backstory: >
    You are pretty good at math

reporting_analyst:
  role: >
    You are an Automated Grading System designed to evaluate mathematical solutions with precision.
  goal: >
    Evaluate the response with precision, comparing it to the correct answer. Determine if the solution is correct.
  output_format: >
    Present your final evaluation as a score of '1' (correct) or '0' (incorrect) only. Do not include any explanatory text or justification.
  backstory: >
    You are a specialized verification algorithm optimized for mathematical equivalence detection across various forms and notations.
        \end{minted}
    \end{minipage}
    \hfill
    \begin{minipage}[t]{0.48\textwidth}
        \centering
        \colorbox{green!45}{\parbox{0.95\textwidth}{\centering\textbf{After}}}
        \begin{minted}[breaklines, breakanywhere, fontsize=\small, frame=leftline, escapeinside=||]{yaml}
researcher:
  role: >
    Math solver.
  goal: >
    Solve the mathematical problem {topic} and provide only the final numerical answer or mathematical expression with minimal essential working steps.
    Avoid explanatory text, introductions, or reflections that don't directly contribute to the solution.
  backstory: >
    You are a mathematics expert capable of solving a wide range of problems.

reporting_analyst:
  role: >
    Automated Grading System that evaluates mathematical solutions for problem {topic} against the correct answer {answer}.
  goal: >
    Verify if the response is mathematically equivalent to the correct answer.
  output_format: >
    Present your final evaluation as a score of '1' (correct) or '0' (incorrect) only.
  backstory: >
    Verification system specialized in detecting mathematical equivalence across different notations and forms.
\end{minted}
\end{minipage}
    \caption{Comparison of original (left) and Artemis-optimized (right) YAML configurations.}
    \label{fig:prompt_comparison}
\end{figure}

\begin{figure}[H]
    \centering
    \begin{minipage}[t]{0.48\textwidth}
        \centering
        \colorbox{red!45}{\parbox{0.95\textwidth}{\centering\textbf{Before}}}
    \begin{minted}[breaklines,breakanywhere,fontsize=\small,frame=leftline,escapeinside=||]{markdown}
    SOLVING = """
    Your task is to solve the math problem step by step and give the final answer as a single number in the simplest way possible. Avoid advanced theories and unrelated examples. 
    
    Return ONLY JSON matching this schema:
    ```json
    {{
      "result": ["Step 1:...", "Step 2:...", "...", "Answer: ..."]
    }}
    ```
    Now, solve this problem based on these context:
    - Math Problem: 
    {problem}
    """
    \end{minted}
    \end{minipage}
    \hfill
    \begin{minipage}[t]{0.48\textwidth}
        \centering
        \colorbox{green!45}{\parbox{0.95\textwidth}{\centering\textbf{After}}}
    \begin{minted}[breaklines,breakanywhere,fontsize=\small,frame=leftline,escapeinside=||]{markdown}
    SOLVING = """
    Your task is to solve the math problem in clear, small steps.
    
    Requirements:
    - Show all intermediate calculations explicitly (write out each arithmetic operation with the exact numbers used).
    - Verify each arithmetic operation before proceeding to the next step by adding a Check line (e.g., "Check: 24 + 16 = 40").
    - Decompose complex problems into smaller sub-problems and solve them sequentially.
    - Do not skip or assume any computational steps; each step must build logically from the previous one with explicit numerical results.
    - Clearly label the final answer on the last line as "Answer: <number>" (a single number in simplest form).
    - Avoid advanced theories and unrelated examples.
    
    Return ONLY JSON matching this schema:
    ```json
    {{
      "result": ["Step 1: ...", "Check: ...", "Step 2: ...", "Check: ...", "...", "Answer: ..."]
    }}
    ```
    Now, solve this problem based on these context:
    - Math Problem: 
    {problem}
    """
    \end{minted}
\end{minipage}
    \caption{Comparison of original (left) and Artemis-optimized (right) prompt configurations.}
    \label{fig:mathtale_prompt_comparison}
\end{figure}

\end{document}